\documentclass{aastex}
\usepackage{emulateapj5}

\shorttitle{M32}
\shortauthors{ALISTER W.\ GRAHAM}
\slugcomment{accepted for publication in ApJ Letters, Feb 2002}

\begin{document}

\title{EVIDENCE FOR AN OUTER DISK IN THE PROTOTYPE `COMPACT ELLIPTICAL' 
GALAXY M32} 
\author{Alister W. Graham\altaffilmark{1}}
\affil{Instituto de Astrof\'{\i}sica de Canarias, La Laguna,
E-38200, Tenerife, Spain}
\altaffiltext{1}{Current address: Department of Astronomy, Univ.\ of Florida, 
Gainesville, FL, USA (Graham@astro.ufl.edu)}

\begin{abstract}
M32 is the prototype for the relatively rare class of galaxies referred 
to as {\it compact ellipticals}.
It has been suggested that M32 may be a tidally disturbed $r^{1/4}$ 
elliptical galaxy, or the remnant bulge of a disk-stripped early-type 
spiral galaxy. 
This paper reveals that the surface brightness profile, the 
velocity dispersion measurements, and the estimated supermassive 
black hole mass in M32 are inconsistent with the galaxy 
having, and probably ever having had, an $r^{1/4}$ light profile. 
Instead, 
the radial surface brightness distribution of M32 resembles an almost 
perfect (bulge + exponential disk) profile, which is accompanied by a 
marked increase in 
the ellipticity profile and an associated change in the position
angle profile where the `disk' starts to dominate.  Compelling 
evidence that this bulge/disk interpretation is accurate comes from 
the best-fitting $r^{1/n}$ 
bulge model which has a S\'ersic index $n=1.5$, in agreement with the
recently discovered relation between a bulge's S\'ersic index and 
the mass of its supermassive black hole.  An index $n\geqslant4$ 
would also be inconsistent with the stellar velocity dispersion of M32. 
The bulge-to-disk size ratio $r_{\rm e}/h$ equals 0.20, and the logarithm  
of the bulge-to-disk luminosity ratio $\log(B/D)$ equals 0.22, typical of 
lenticular galaxies. 
The effective radius of the bulge is 27$\arcsec$ ($\sim$100 pc), while 
the scale-length of the 
disk is less well determined: due to possible tidal-stripping of the 
outer profile beyond 220-250$\arcsec$, 
%
%
the scale-length may be as large as 1.3 kpc.  M32 is a relatively 
face-on, nucleated, dwarf galaxy with a low surface brightness disk
and a high surface brightness bulge.  This finding brings into question 
the very existence of the compact elliptical class of galaxies. 
\end{abstract}


\section{Introduction}

The 32nd object in the catalog of Messier (1850) has come to be known as 
the archetype of high surface brightness, low-luminosity, {\it compact 
elliptical} galaxies (de Vaucouleurs 1961).  
It has been proposed that they may be the dense cores of tidally truncated,
or at least modified, 
ordinary elliptical galaxies (King 1962; Faber 1973; 
Nieto \& Prugniel 1987; Choi, Guhathakurta, \& 
Johnston 2002, and references therein).  
It has also been suggested that M32 (NGC~221) may 
in fact never have been an elliptical galaxy, but is instead 
the bulge of a (partially) stripped disk galaxy (Bekki et al.\ 
2001; see also Nieto 1990). 

Using the tight ($r_s=0.91$) correlation between the central concentration 
index $C_{r_{\rm e}}(1/3)$ of a bulge\footnote{By the term bulge 
it is meant both an elliptical galaxy and the bulge of a disk galaxy.}
and the mass of its central supermassive black hole (SMBH; 
Graham et al.\ 2001a), provides a new constraint capable of determining 
which proposition, if either, is correct.  
Simply by modelling M32's surface brightness profile as either a
one-component ($r^{1/n}$) elliptical, or as a 2-component 
($r^{1/n}$ bulge + exponential) lenticular or spiral galaxy, and knowing its 
SMBH mass, allows one to decipher which scenario is more probable. 

The paper is laid out as follows. 
Section 2 introduces M32's surface brightness profile which 
is modelled in Section 2.1.  Section 3 introduces M32's 
velocity dispersion measurements and SMBH mass estimates into 
the discussion, supporting the notion that the bulge 
{\it and disk} decomposition is the correct one.  

\section{The surface brightness profile of M32}

M32's major-axis, $R$-band surface brightness profile presented in 
Kent (1987) is reanalyzed here.  The data was obtained using an RCA CCD 
attached, at different times, to three telescopes with different 
fields of view at the Whipple Observatory on Mount Hopkins. 
The observing procedures are described in Kent (1983) and the 
(remarkably standard) reduction procedure given in Kent (1987). 

The main complication with this galaxy is its proximity to M31 (Andromeda). 
It resides (in projection) within the outer disk of M31, and so the
disk of M31 had to first be modelled and then subtracted.  Kent 
wrote, ``{\it The light from M31 was removed 
approximately by fitting second order polynomials to the 
background light in the frame of M32 obtained with the 
Bausch \& Lomb 8000} [a 20 cm aperture telescope with a $15\arcmin 
\times 25\arcmin$ field of view] {\it excluding an area about M32 
itself}''. 
Ultimately, uncertainties in the sky-background level resulted in the 
termination of the profile at 
$\mu(r=280$\arcsec$)\sim$24 $R$-mag arcsec$^{-2}$.

Further complications and solutions are described in Kent (1987), 
and won't be repeated here for the following reason. 
After commencing this work, Choi et al.\ (2002) 
presented new
$B$- and $I$-band surface brightness profiles for M32.  These 
show very good agreement over the radial range in common
with Kent (1987), and extend to 420$\arcsec$.  
This lends confidence that the publicly available data 
from Kent (1987) is reliable. 
The profiles of Choi et al.\ (2002) will however be referred to 
in a qualitative manner.

\subsection{Modelling the surface brightness profile}

Inner components such as nuclear disks, star clusters, 
flattened cores, etc., are known to reside within the central $\sim1\arcsec$ 
of many galaxy's (e.g., Rest et al.\ 2001; Ravindranath et al.\ 2001). 
We wish to avoid such features here as we are presently only concerned 
with the bulge (and outer disk, if one exists) of M32. 
However, typically observed galaxies are considerably more distant than M32. 
At a distance of 0.8 Mpc, with the exception of the Milky Way, 
M32 is some 10 times closer than
any other galaxy with a positive SMBH detection 
(Kormendy \& Gebhardt 2001; Merritt \& Ferrarese 2001b). 
Its nuclear region is therefore very well resolved: 
1$\arcsec$ is equivalent to 3.87 pc. 
If M32 was located at the distance of the Virgo cluster, exclusion of its 
central arcsecond would translate to the removal of the inner 
$\sim$ 60-80 pc ($\sim$ 15-20$\arcsec$ at its present location). 

B\"oker et al.\ (2001) chose to model 
the central excesses (above that of the bulge) found in HST images 
of spiral galaxies with a power-law.  
Tonry (1984) noted that the inner 10$\arcsec$ of M32 displays a 
power-law shape.  This may therefore
be a somewhat common feature of bulges\footnote{In passing, it is 
noted that this may be a feature that 
results in the over-estimation of the S\'ersic index $n$ when 
modelling ground-based images which have not resolved, or 
avoided, such central excesses.}. 
Michard \& Nieto (1991) presented evidence that the inner $5\arcsec$ of M32 
contains a nuclear disk, however this has subsequently been refuted 
by Lauer et al.\ (1998) who found no photometric evidence for 
separate nuclear components in HST images. 
However, the presence of an apparent excess central flux did result in 
Kent (1987) excluding the inner 15$\arcsec$ from his model fitting, and 
Choi et al (2002) excluding the inner 10$\arcsec$, 
which is also done here.  Apart from this, the entire
surface brightness profile beyond 10$\arcsec$ is modelled here. 
The data from Table 3 of Kent (1987), and also a resampling of Kent's 
Figure 1 (to give an equal spacing in radius, effectively providing a 
more even weighting [radially] to the data), are modelled. 

\begin{figure}
\epsscale{0.75}
\plottwo{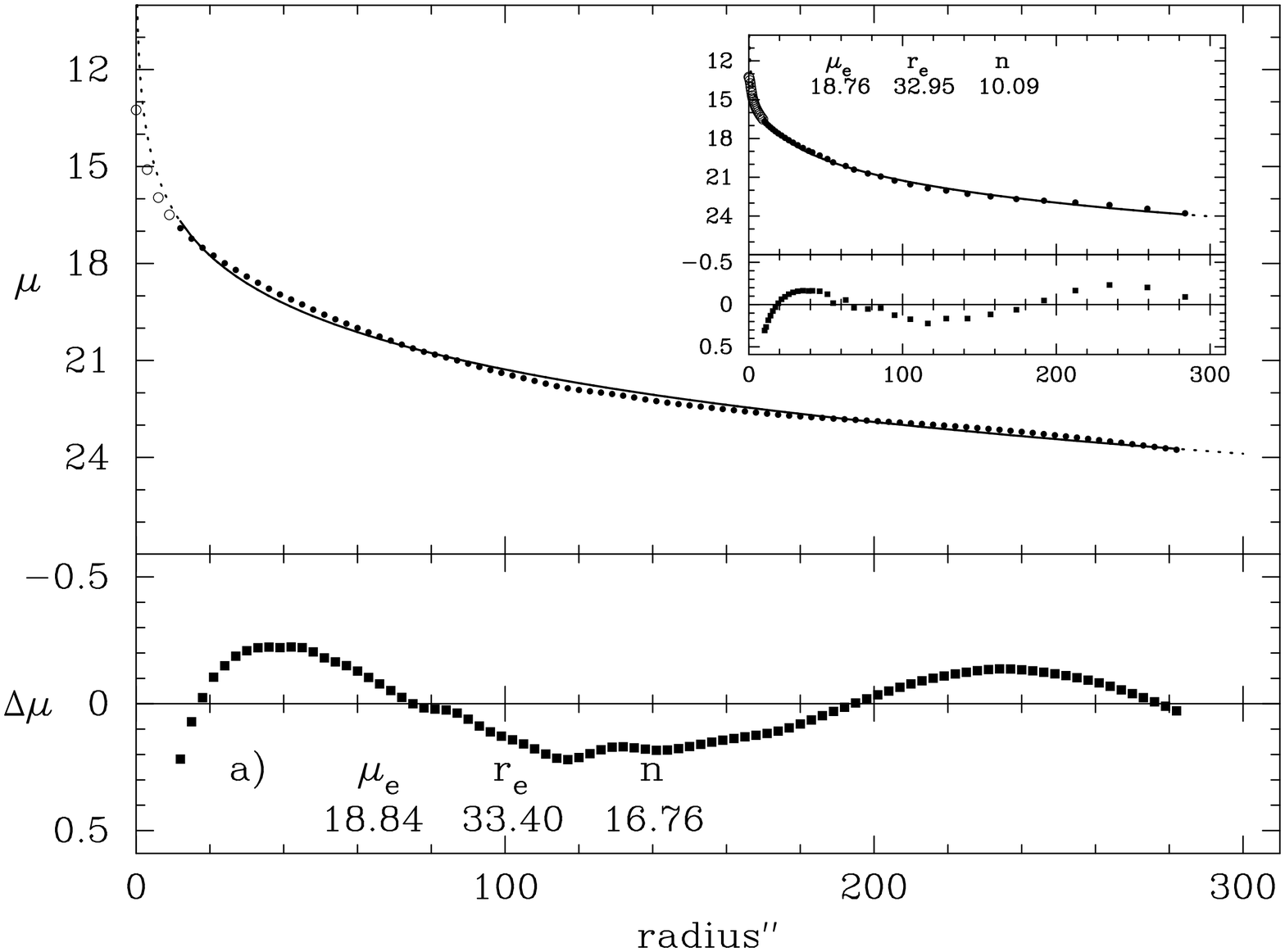}{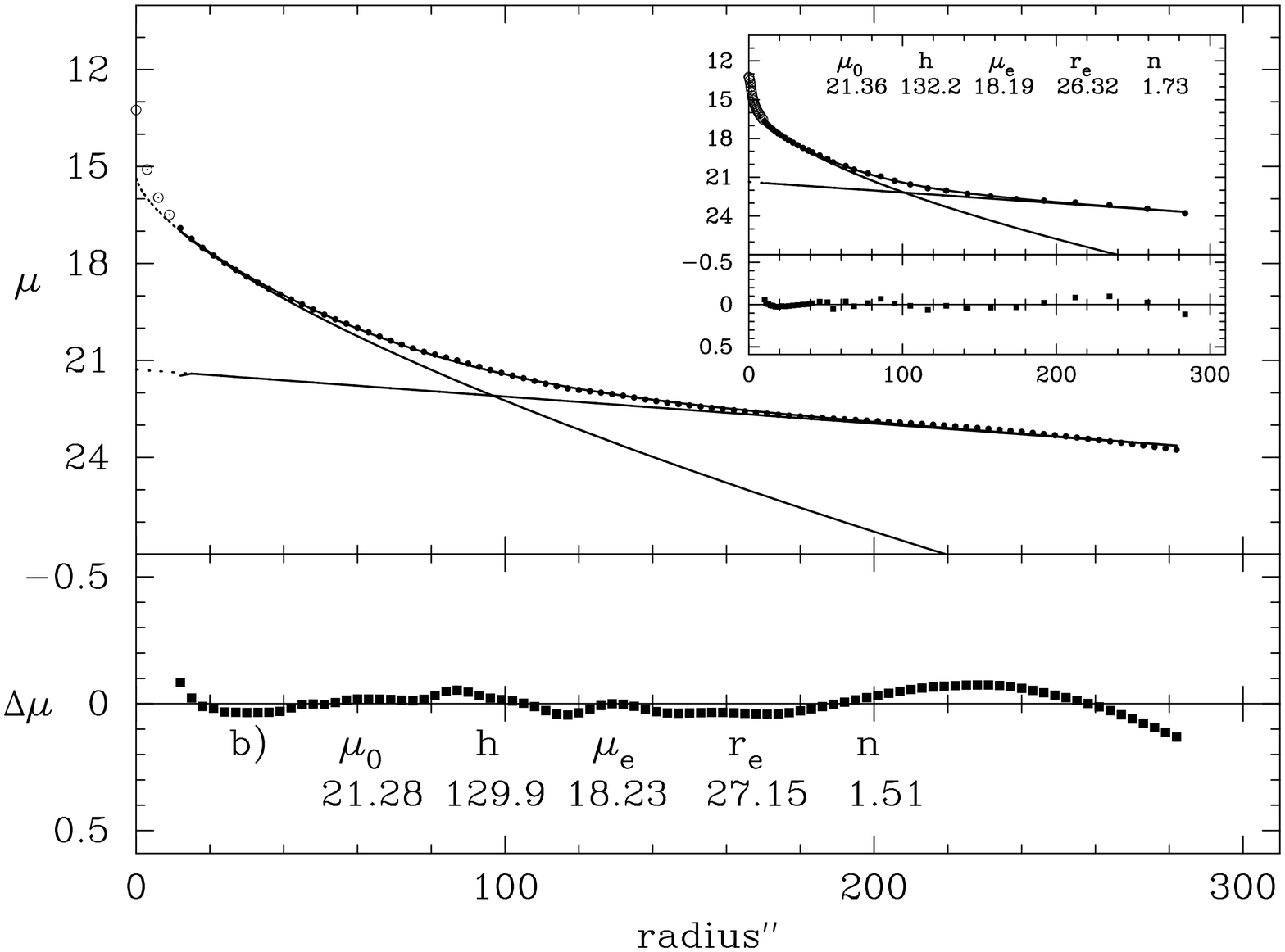}
\caption{M32's (major-axis) $R$-band surface brightness profile 
from Kent (1987), resampled at equal spacing in radius, 
is modelled with 
a) a seeing-convolved $r^{1/n}$-only model, and 
b) a seeing-convolved ($r^{1/n}$ + exponential) model.  
Following common practice, the inner 
10$\arcsec$ have been excluded from the fit. 
The inset figures show the results using the logarithmically spaced 
data from Kent (1987). 
}
\label{fig1}
\end{figure}

Both a seeing-convolved $r^{1/n}$ model and a seeing-convolved 
($r^{1/n}$ + exponential) model were fitted.  
In both cases, all model parameters were simultaneously fitted 
using the quasi-Newton, non-linear least-squares algorithm UNCMND
(Kahaner, Moler, \& Nash 1988) which was 
iterated until convergence on the optimal solution giving the 
smallest $\chi ^2$ value.
All parameters were allowed to range freely in the fitting process, 
the only constraint was that they must be positive real numbers. 
The seeing 
was reported by Kent (1987) to have a FWHM of 1.3$\arcsec$, and
therefore, due to the exclusion of the inner 10$\arcsec$, has 
little effect on the results which are shown in Figure~\ref{fig1}.  
Several truncations of the outer profile were explored, but 
the overall conclusion was always the same: 
{\it M32 cannot be modelled as, that is to say it is not 
(structurally), a single component system}.  The curvature in the
residual profile of Figure~\ref{fig1}a is classic 
evidence of this. 
Importantly, M32 can be described exceptionally well with the standard 
two-component model used these days to model disk galaxies with 
bulges\footnote{Fitting an ($r^{1/4}$ + exponential) model resulted in 86\% 
more scatter than using an ($r^{1/n}$ + exponential) model, and a rather
unimpressive fit.}.
%
%
The notably small residuals are compelling evidence that this model 
is likely to be correct.  This is, however, not to say that M32 has 
a rotationally supported disk of stars, only that is has an outer exponential
distribution of stars.  To help evaluate if the two-component model 
is indeed correct, let us look at the resulting structural parameters 
to see if they are consistent with those of known bulge/disk systems. 

The bulge-to-disk size ratio ($r_{\rm e}/h$) is 0.20, in good
agreement with that of normal disk galaxies (Graham 2001, and references
therein) and suggests nothing unusual.  
The logarithm of the bulge-to-disk luminosity ratio ($\log B/D$)
is 0.22 and typical of an S0 galaxy.  
The effective half-light radius of the bulge is 27$\arcsec$, 
in reasonable agreement with the value of 32$\arcsec$ derived by Kent
when modelling the radial interval $15\arcsec<r<100\arcsec$, 
and even agrees well with the value of 
30$\arcsec$ found by de Vaucouleurs (1953). 
The effective bulge surface brightness is 18.23 $R$-mag arcsec$^{-2}$, 
the central bulge surface brightness is 15.31 $R$-mag arcsec$^{-2}$, 
and the absolute magnitude of the bulge is 16.34 $R$-mag.

Further support for the above bulge/disk decomposition comes from the new 
location of M32 in several structural parameter diagrams for bulges. 
M32 is a distant outlier in the insightful $\log n$-$B_T$ diagram presented in 
Jerjen \& Binggeli (1997; their figure 2).  
Noting that their value of $n$ corresponds to our value of $1/n$, 
when $n=1.5$ (in our notation) M32
moves up into the very center of points, and indeed the very center of
the relation defined by the dwarf galaxies (see also 
Graham 2001, his figure 14).  The reason for this is that the 
disk had biased Jerjen \& Binggeli's one-component $r^{1/n}$ 
fit to M32's light profile, resulting in a value of $n$ which 
is $\sim$5 in our notation. 
This additionally explains the deviant nature of M32 in the 
$\log r_0$-$B_T$ diagram, and most of the discrepancy in the  
$\mu_{\rm 0,bulge}$-$B_T$ diagram.   
The suggestion by Wirth \& Gallagher (1984) that compact elliptical
galaxies may be the extension of brighter elliptical galaxies, 
based on color, luminosity, and velocity dispersion (and the 
existence of isolated compact elliptical galaxy candidates) 
would appear to be correct in the sense that the bulge of M32 
appears to be the faint extension of bulges and elliptical 
galaxies in general.  

What of the disk? 
Kent (1987) remarked that there ``{\it seems to be an excess of light
at large radii, with the excess having an exponential profile}''. 
The ellipticity profile of Kent (1987) also suggests the 
presence of a distinct outer component in M32, rising from 
$\epsilon$=0.11 at 150$\arcsec$ to $\epsilon$=0.19 by $\sim200\arcsec$
(where the ellipticity was then held constant). 
This feature is even more clearly evident in the ellipticity profile 
of Choi et al.\ (2002), rising steadily from $\epsilon\sim0.14$ 
at 100$\arcsec$ to $\sim0.35$ at 250$\arcsec$.  
The position angle also changes notably at 100$\arcsec$-150$\arcsec$.
%

The central disk surface brightness is 21.28 $R$-mag arcsec$^{-2}$ 
(Figure~\ref{fig1}b). 
Applying the standard disk inclination correction $-2.5C\log(1-\epsilon)$,
where $\epsilon$ is the ellipticity of the disk ($\epsilon\sim$0.3 from
Choi et al.\ 2002) and $C$=0.5 in the $R$-band (Tully \& Verheijen 1997), 
would give a face-on central disk surface brightness of 21.48 
$R$-mag arcsec$^{-2}$.  
The definition of a low surface brightness (LSB) galaxy, 
is one in which the central disk surface brightness is more than 
one magnitude fainter than the canonical Freeman (1970) value of 
21.65 $B$-mag arcsec$^{-2}$ (which is about 20.5 $R$-mag arcsec$^{-2}$). 
%
%
Thus, M32 would just about qualify as a bulge-dominated LSB disk galaxy
(Beijersbergen, de Blok, \& van der Hulst 1999; Impey \& Bothun 1997). 
It is, however, because of it's size and magnitude, a dwarf galaxy. 

\begin{figure}
\epsscale{0.75}
\plotone{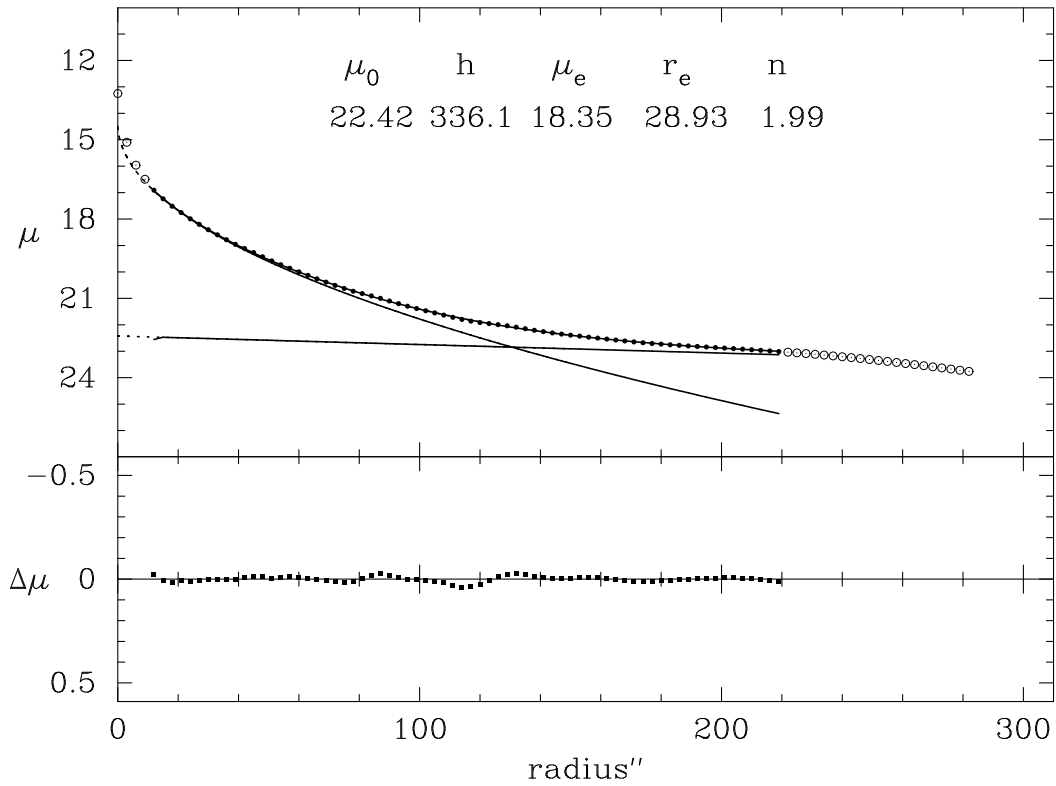}
\caption{M32's (major-axis) $R$-band surface brightness profile 
from Kent (1987) is modelled here with a seeing-convolved 
($r^{1/n}$ + exponential) model. 
The inner 10$\arcsec$ have been excluded from the fit, as has the 
data beyond 220$\arcsec$. 
}
\label{fig2}
\end{figure}

The scale-length of the disk is 130$\arcsec$ or 0.5 kpc. 
There is, however, possible evidence of a disturbance in the very 
outer profile; 
it turns downward from an exponential profile at around 250$\arcsec$.  
That is, there appears to be a lack of light, relative to the 
exponential part of the profile, beyond $\sim250\arcsec$.  
This behavior is visible in both the data of Kent (1987) and also that of 
Choi et al.\ (2002).  It is accompanied 
by a marked change in the behavior of the ellipticity profile of Choi 
et al.\ (2002), flattening (or even decreasing slightly) beyond 
250$\arcsec$.  This may be a sign that 
material has been stripped away from the outer disk, although it is 
stressed that this conclusion is largely speculative. 
However, if true, this turnover would be biasing the surface brightness 
profile fit, making the disk scale-length appear shorter than it 
was before tidal-stripping commenced. 
Truncating the (equally spaced) profile at 220$\arcsec$, to avoid the 
potentially stripped outer disk, and thereby (possibly) sampling only the 
original undisturbed 
profile gives $h=336\arcsec$ (1.27 kpc) and $\mu_0=22.42$ for the disk. 
For the bulge, $r_{\rm e}=29\arcsec$ and $n=1.99$ 
(see Figure~\ref{fig2}; using the 
profile which is logarithmically spaced in radius gives $n=2.08$). 
Here the value of $\log B/D$=-0.09, and 
$r_{\rm e}/h$=0.09, which are again not unreasonable. 
An $r^{1/n}$-only model fails to provide a convincing fit to this
truncated radial range. 


\section{Discussion}

The surface brightness profile of M32 can be modelled remarkably well as 
a combination of an $r^{1/n}$ bulge and an exponential disk of stars. 
The ellipticity profile additionally supports this interpretation of the
data.  However, could, instead, M32's surface brightness profile be so 
disturbed that the outer exponential envelope is actually due to material 
pulled off from what was once a one-component $r^{1/4}$ (i.e.\ $n=4$) 
elliptical galaxy?   This scenario appears unlikely for the following 
reasons. 

Firstly, somewhat persuasive evidence comes from the 
the central velocity dispersion of M32 which has been measured to be
76$\pm$10 km s$^{-1}$ (van der Marel et al.\ 1998). 
This figure is in good agreement with the average value of 
74 km s$^{-1}$ obtained from numerous estimates listed in 
Hypercat.\footnote{Hypercat can be reached at
\url{http://www-obs.univ-lyon1.fr/hypercat/}}  
It also agrees with Gebhardt et al.'s (2000) estimate of 75 km s$^{-1}$ 
for the luminosity weighted velocity dispersion within one effective radius. 
%
Recently, it has been discovered that 
the central velocity dispersion of a bulge correlates
strongly ($r=0.8$) with the shape of the bulge light profile
(as measured with the S\'ersic index: Graham et al.\ 2001b; Graham 2002).  
Galaxies with measured velocity 
dispersions less than 100 km s$^{-1}$ are observed to have values 
of $1<n<2$.  The measured value of $n=1.5$ for the bulge of M32 is thus
exactly what one would expect from its dynamics, not $n=4$. 
Bulges (which includes elliptical galaxies) with values of
$n\gtrsim4$ have velocity dispersions typically greater than 
100 km s$^{-1}$. 

The second, related, line of reasoning comes from M32's SMBH mass. 
From ground based kinematical data, Tonry (1984,7) predicted 
a SMBH mass of (3-10)$\times10^6 M_{\sun}$ at the center of M32. 
This pioneering work has been confirmed with HST data, from which 
van der Marel et al.\ (1998) derived a SMBH mass of 
(3.4$\pm$0.7)$\times10^6 M_{\sun}$, and more recently 
Joseph et al.\ (2001) find a mass of (2-4)$\times10^6 M_{\sun}$. 
%
%
The velocity dispersion ($\sigma$) and SMBH mass ($M_{\rm bh}$) 
combination of M32 is known to fall on the 
$\log M_{\rm bh}$-$\log\sigma$ relation for (non-disturbed) ellipticals
and bulges (Ferrarese \& Merritt 2000; Gebhardt et al.\ 2000; 
Merritt \& Ferrarese 2001a).  Thus, unless the process of tidal stripping
modifies both the central velocity dispersion and the SMBH mass in such a 
way that it preserves the $\log M_{\rm bh}$-$\log\sigma$ relation, 
the central structure and dynamics of M32 are likely to be that of the 
original (i.e.\ undisturbed) galaxy. 
Indeed, due to the higher densities at smaller radii, 
the central velocity dispersion and 
hence inner mass distribution and therefore inner light profile 
are expected to remain largely unaffected by the outer stripping process. 
Taken with the result in the previous paragraph, this strongly suggests 
that the $n=1.5-2.0$ bulge profile is the original shape of the bulge. 

\begin{figure}
\epsscale{0.75}
\plotone{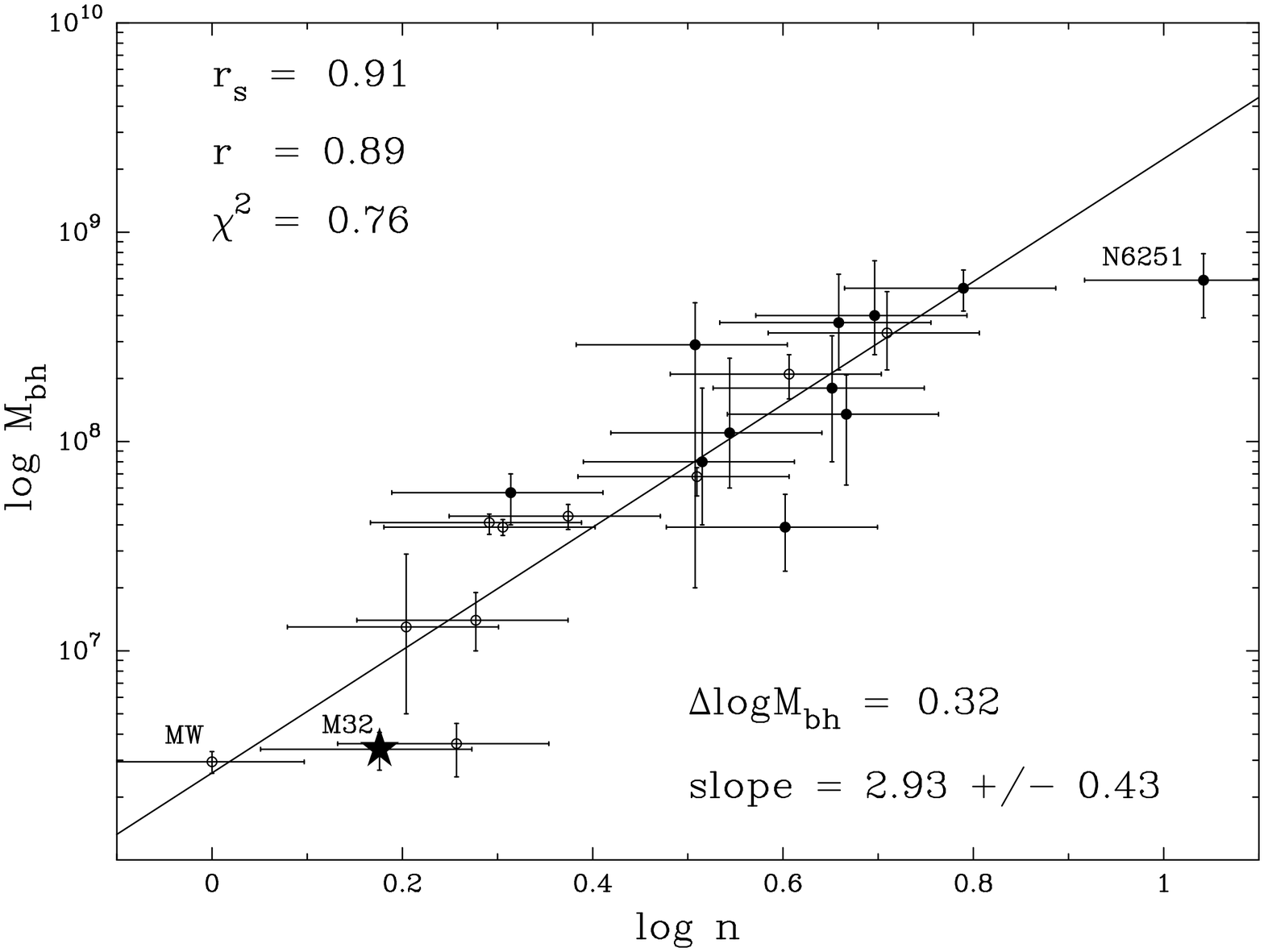}
\caption{The location of M32 is shown in the $\log M_{\rm bh}-\log n$ 
diagram by the star.  The value of $n=1.5$ comes from the 
($r^{1/n}$ bulge + exponential) model (Figure~\ref{fig1}c).  
Following Graham et al.\ 2001b, a typical error of 25\% for $n$ is shown. 
The SMBH mass estimate comes from van der Marel et al.\ (1998). 
The regression and statistics have been performed excluding M32,
so it in no way biases the fit. 
}
\label{fig3}
\end{figure}

Figure~\ref{fig3} shows the $\log M_{\rm bh}$-$\log n$ relation 
for bulges.  It is a variant of the relation between SMBH mass and 
central bulge concentration ($C_{r_{\rm e}}[1/3]$) shown in 
Graham et al.\ (2001a).  (The parameter $C_{r_{\rm e}}(1/3)$ is a 
monotonicly increasing function of $n$, Trujillo et al.\ 2001.) 
If M32 did ever have an $r^{1/4}$ profile, then (from Figure~\ref{fig3}) 
its SMBH mass should have once been and should still be 
$\sim10^8 M_{\sun}$, and certainly greater than $\sim10^7 M_{\sun}$.  
Given the bulge value of $n=1.5$ agrees with the actual SMBH mass estimate
(and with M32's velocity dispersion), the stars composing the outer 
exponential envelope are almost certainly an excess (relative to the bulge) 
which have not come from a reshaped bulge\footnote{The reason Choi et al.\
(2002) claimed to be able to fit an $r^{1/4}$ bulge to the inner 
profile of M32 is likely because of the limited radial range 
$10\arcsec<r<30\arcsec$ they used to do this.}. 
%
%

All of this is not
to say that the outer profile of M32 has not been disrupted,  Indeed,
the deficit of stars in the outer profile, causing the downward kink 
(or break) in the disk, may have been due to gravitational 
stripping by M31.  This deficit is also visible in the $B$- and 
$I$-band profiles of Choi et al.\ (2002) and hence less likely to
be a systematic error in the profile extraction technique of Kent (1987). 
Indeed, the recently reported tidal stream of metal-rich stars around M31 
is thought to have likely come from M32 and/or NGC~205 (Ibata et al.\ 2001).  
The presence of a disk may also explain the intermediate-age (5-15 Gyr) 
stellar population in M32 
(e.g.\ Grillmair et al.\ 1996; Davidge 2000). 




Perhaps a comment on nomenclature would be beneficial here. 
Kormendy \& Gebhardt (2001) used the words `compact' and `fluffy' 
when referring to bulges.  It should be noted that this has no 
reference at all to varying profile `shape' (or equivalently 
`concentration', 
according to the mathematical definition given in Trujillo et al.\ 2001, 
their equation 5 and 6). 
That is, no reference to departures from the $r^{1/4}$ law, or 
structural homology, are implied by their terms `compact' and `fluffy'.  
The $r^{1/4}$ law only has a horizontal scale term ($r_{\rm e}$) and a
vertical scale term ($\mu_{\rm e}$ or $\mu_0$), the shape, or 
concentration, is exactly the same for all $r^{1/4}$ profiles. 
Kormendy \& Gebhardt (2001) discussed variations in $\mu_e$ and $r_e$
when they referred to `compact' and `fluffy'. 
Hence, although M32 is regarded as compact, it's
central concentration $C_{r_{\rm e}}(1/3)$ is actually rather low. 

Bekki et al.'s (2001) N-body/SPH simulations of tidal interactions between 
M31 and an orbiting early-type spiral galaxy predict 
either a complete stripping of the disk, or at least a vertical heating of 
the satellite's disk to create a thick disk.  
Clearly a faint disk, with very little gas (Welch \& Sage 2001), still 
surrounds M32.  Much of the gas and stars may indeed have been stripped 
away, resulting in the low surface brightness disk. 
Also possible, is the suggestion by Bekki et al.\ (2001) that tidal 
interactions with M31 funnelled some of M32's gas to its center, 
forming a massive 
starburst (see also Noguchi \& Ishibashi 1986).  This could 
account for the excess central flux within the inner 
$\sim10\arcsec$ of M32 having an age of $\sim$4 Gyrs 
(Vazdekis \& Arimoto 1999; del Burgo et al.\ 2001).




%

Compact elliptical galaxies are a rare class of objects. 
A closer inspection of such objects 
%
%
seems warranted in order to inspect whether the species is indeed real, 
or simply a case of misclassification.

%
%
%

\acknowledgements
I wish to thank Peter Erwin for providing me with Kent's (1987) 
surface brightness profile of M32, resampled with equal 
spacing in radius, and for useful discussions which helped to 
shape this paper.  
I am also grateful to Carme Gallart for her comments on this work.

\clearpage

\begin{references}
\reference{Bdv99}Beijersbergen, M., de Blok, W.J.G., \& van der Hulst, J.M.\ 1999, A\&A, 351, 903
\reference{Bek01}Bekki, K., Couch, W.J., Drinkwater, M.J., Gregg, M.D., 2001, ApJ, 557, L39
\reference{Bok01}B\"oker T., Laine, S., van der Marel, R.P., Sarzi, M., Rix, H.-W., Ho, L.C., Shields, J.C., 2001, AJ, submitted (astro-ph/0112086)
\reference{Cet01}Choi, P.L., Guhathakurta, P., Johnston, K.V., 2002, AJ, submitted (astro-ph/0111465)
\reference{Dav00}Davidge, T.J., 2000, PASP, 112, 1177
\reference{del01}del Burgo, C., Peletier, R.F., Vazdekis, A., Arribas, S., Mediavilla, E., 2001, MNRAS, 321, 227
\reference{deV53}de Vaucouleurs, G., 1953, MNRAS, 113, 134
\reference{deV61}de Vaucouleurs, G., 1961, ApJS, 5, 233 
\reference{Fab73}Faber, S.M., 1973, ApJ, 179, 423 
\reference{FaM00}Ferrarese, L., \& Merritt, D.\ 2000, ApJ, 539, L9
\reference{Fre70}Freeman, K.C.\ 1970, \apj, 160, 811
\reference{Get00}Gebhardt, K., et al.\ 2000, ApJ, 539, L13
\reference{Gra01}Graham, A.W.\ 2001, AJ, 121, 820
\reference{Gra02}Graham, A.W.\ 2002, MNRAS, submitted
\reference{Get01}Graham, A.W., Erwin, P., Caon, N., \& Trujillo, I., 2001a, ApJ, 563, L11
\reference{GTC01}Graham, A.W., Trujillo, I., \& Caon, N.\ 2001b, AJ, 122, 1707
\reference{Gri96}Grillmair, C.J., et al., 1996, AJ, 112, 1975
\reference{Iet01}Ibata, R., Irwin, M., Lewis, G., Ferguson, A.M.N., Tanvir, N., 2001, Nature, 412, 49
\reference{IaB97}Impey, C., \& Bothun, G., 1997, ARA\&A, 35, 267
\reference{JaB97}Jerjen, H., \& Binggeli, B.\ 1997, in The Nature of Elliptical Galaxies; The Second Stromlo Symposium, ASP Conf.\ Ser., 116, 239
\reference{Jet01}Joseph, C.L., et al., 2001, ApJ, 550, 668
\reference{KMN88}Kahaner, D., Moler, C., \& Nash, S.G., 1988, ``Numerical Methods and Software, Prentice Hall
\reference{Ket83}Kent, S.M., 1983, ApJ, 266, 562
\reference{Ket87}Kent, S.M., 1987, AJ, 94, 306
\reference{Kin62}King, I.R., 1962, AJ, 67, 471
\reference{KaG01}Kormendy, J., \& Gebhardt, K.\ 2001, in The 20th Texas Symposium on Relativistic Astrophysics, ed.\ H.\ Martel, \& J.C.\ Wheeler, AIP, in press (astro-ph/0105230)
\reference{Lau98}Lauer, T.R., Faber, S.M., Ajhar, E.A., Grillmair, C.J., Scowen, P.A., 1998, AJ, 116, 2263
\reference{MaF01a}Merritt, D., \& Ferrarese, L.\ 2001a, ApJ, 547, 140
\reference{MaF01b}Merritt, D., \& Ferrarese, L.\ 2001b, in ASP Conf.\ Ser.\ 249, The Central Kpc of Starbursts and AGN: the La Palma Connection, eds.\ J.H.\ Knapen, J.E.\ Beckman, I.\ Shlosman \& T.J.\ Mahoney, (San Francisco: ASP), 335
\reference{M1850}Messier, C., 1850, Connaissance des Temps, 1784, 227-269
\reference{MaN91}Michard, R., \& Nieto, J.-L., 1991, A\&A, 243, L17
\reference{Nie90}Nieto, J.-L., 1990, in Dynamics and Interactions of Galaxies, ed.\ R.\ Wielen (Berlin: Springer), 258
\reference{NaP}Nieto, J.-L., \& Prugniel, P., 1987, A\&A, 186, 30
\reference{NaI86}Noguchi, M., \& Ishibashi, S., 1986, MNRAS, 219, 305
\reference{Rav01}Ravindranath, S., Ho, L.C., Peng, C.Y., Filippenko, A.V., Sargent, W.L.W.\ 2001, AJ, 122, 653
\reference{Ret01}Rest, A., et al.\ 2001, AJ, 121, 2431
\reference{Ser68}S\'ersic, J.-L.\ 1968, Atlas de Galaxias Australes (Cordoba: Observatorio Astronomico)
\reference{Ton84}Tonry, J., 1984, ApJ, 283, L27
\reference{Ton87}Tonry, J., 1987, ApJ, 322, 632
\reference{TGC01b}Trujillo, I., Graham, A.W., \& Caon, N.\ 2001, MNRAS, 326, 869
\reference{TaV97}Tully R.B., Verheijen M.A.W., 1997, ApJ, 484, 145
\reference{vet98}van der Marel, R.P., Cretton, N., de Zeeuw, P.T., Rix., H.-W., 1998, ApJ, 493, 613
\reference{ZaA99}Vazdekis, A., \& Arimoto, N., 1999, ApJ, 525, 144
\reference{WaG84}Wirth, A., \& Gallagher, J.S., III, 1984, ApJ, 282, 85
\reference{WaS01}Welch, G.A., \& Sage, L.J., 2001, ApJ, 557, 671
\end{references}
\end{document}